\documentclass[sigconf]{acmart}
\AtBeginDocument{%
  }

\setcopyright{acmlicensed}
\copyrightyear{2018}
\acmYear{2018}
\acmDOI{XXXXXXX.XXXXXXX}
\acmConference[XAIxArts 2025]{Explainable AI for the Arts Workshop 2025}{June 23, 2025}{Online}
\acmISBN{978-1-4503-XXXX-X/2018/06}

\usepackage{graphicx}

\begin{document}

\title{DeformTune: A Deformable XAI Music Prototype for Non-Musicians}

\author{Ziqing Xu}
\email{z.xu0520221@arts.ac.uk}
\orcid{0009-0006-9627-201X}
\affiliation{%
  \institution{Creative Computing Institute, University of the Arts London }
  \city{London}
  \country{United Kingdom}
}

\author{Nick Bryan-Kinns}
\email{n.bryankinns@arts.ac.uk}
\orcid{0000-0002-1382-2914}
\affiliation{%
  \institution{Creative Computing Institute, University of the Arts London}
  \city{London}
  \country{United Kingdom}}

\renewcommand{\shortauthors}{Xu and Bryan-Kinns}

\begin{abstract}
Many existing AI music generation tools rely on text prompts, complex interfaces, or instrument-like controls, which may require musical or technical knowledge that non-musicians do not possess. This paper introduces DeformTune, a prototype system that combines a tactile deformable interface with the MeasureVAE model to explore more intuitive, embodied, and explainable AI interaction. We conducted a preliminary study with 11 adult participants without formal musical training to investigate their experience with AI-assisted music creation. Thematic analysis of their feedback revealed recurring challenges—including unclear control mappings, limited expressive range, and the need for guidance throughout use. We discuss several design opportunities for enhancing explainability of AI, including multimodal feedback and progressive interaction support. These findings contribute early insights toward making AI music systems more explainable and empowering for novice users.
\end{abstract}


\maketitle
\section{Introduction and Background}

Recent advances in AI have enabled powerful music generation tools, from text-based systems such as Suno AI  \cite{nugroho_use_2024} 
to intelligent performance systems.  
However, many of these rely on complex interfaces, text based prompting, or musical knowledge \cite{han_understanding_2024}. 
While text prompts can simplify interaction in certain contexts, they often require users to articulate specific musical ideas clearly in language, which can be challenging for novices without prior musical or technical knowledge. Similarly, instrument-like or complex graphical interfaces typically demand existing musical proficiency, thus potentially excluding non-musicians. 
The lack of explainability, making it difficult for users to understand how the system works or influence its creative decisions—an issue that explainable AI (XAI) seeks to address.
Some recent research has explored XAI for music, such as interpretable latent models \cite{bryan-kinns_exploring_2023} and context-aware explanations \cite{privato_context-sensitive_2023}. Yet, these efforts typically target professional musicians or technically skilled users, leaving a gap in support for novices.
Recent work on interactive music systems has explored alternative interface paradigms, such as deformable controllers that afford intuitive, embodied interaction. For example, Zheng and Bryan-Kinns \cite{zheng_squeeze_2022} highlight how deformable materials enable unexpected affordances in musical interaction design, while \citet{wu_soundmorphtpu_2024} demonstrate how gesture mapping through soft TPU-based surfaces can enhance physical engagement.

To address this gap, our system, DeformTune, introduces a deformable tactile interface designed specifically to lower these interaction barriers, providing embodied and intuitive controls to empower non-expert users in music co-creation. This work investigates how explainability and interaction design can better support non-musicians in AI-assisted music creation. Drawing on alternative interface paradigms we explore whether deformable interfaces with their intuitive, embodied control offer opportunities for more transparent interaction, helping non-experts to understand and influence generative music systems without requiring technical knowledge. 
Given the challenges of designing XAI systems for music and the opportunities for deformable interfaces in this area the research questions in this paper are:

\textbf{RQ1:} What interaction challenges do users without musical or technical backgrounds face when engaging in AI-assisted music creation through a deformable interface?

\textbf{RQ2:} What specific explainability (XAI) needs do these users have during co-creation with the system, and in what forms do they prefer to receive explanations?


\section{Method}
To explore non-experts’ explainability needs in AI-assisted music generation, we conducted a user study with 11 adults, aged between 18 and 31, who were all students with no formal musical training, using a custom-built deformable interface referred to as \emph{DeformTune}. Participants were recruited from the author's institution and voluntarily took part. The study had ethical clearance from the institution's research committee. Each 40-minute session included a demographic survey, a 5-minute free exploration, a 4-minute creative task (composing a 10-second ringtone), and two questionnaires to assess user perceptions - UEQ-short \cite{schrepp_design_2017} and RiCE \cite{ford_towards_2023}. This was followed by a semi-structured interview to gather deeper insights. We analyzed the qualitative and quantitative data to identify user challenges and opportunities for improving explainability in AI music making. For the interview we adopted the Question-Driven Design Process proposed by \citet{liao_question-driven_2021}. 
Our interview questions covered core XAI needs, including:
\begin{itemize}
    \item \textbf{Why}: e.g., “Would you like to know why it generated this music?” 
    \item \textbf{How}: e.g., “What do you think the system is doing?”
    \item \textbf{What if}: e.g., “What would help you understand what changed the output?”
    \item \textbf{Data}: e.g., “Do you care about what kind of data trained this model?”
    \item \textbf{Performance}: e.g., “When do you think this system might make a mistake?” 
\end{itemize}

\begin{figure}[htbp]
    \begin{minipage}[t]{0.5\linewidth}
        \centering
        \includegraphics[width=0.6\textwidth]{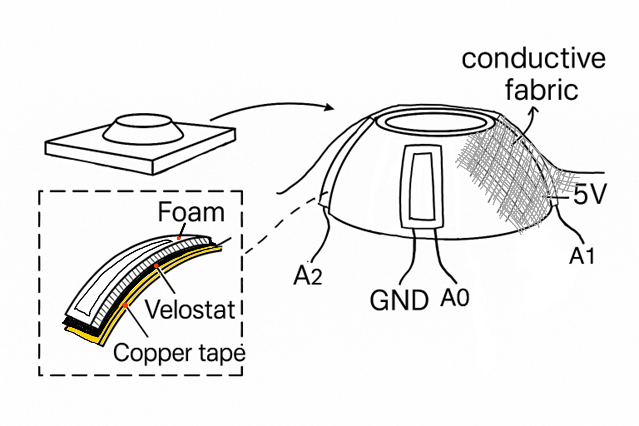}
        \caption{DeformTune Structure.}\label{fig:deformtune_structure}
        \Description{To show the structure of DeformTune.}
    \end{minipage}%
    \begin{minipage}[t]{0.5\linewidth}
        \centering
        \includegraphics[width=0.6\textwidth]{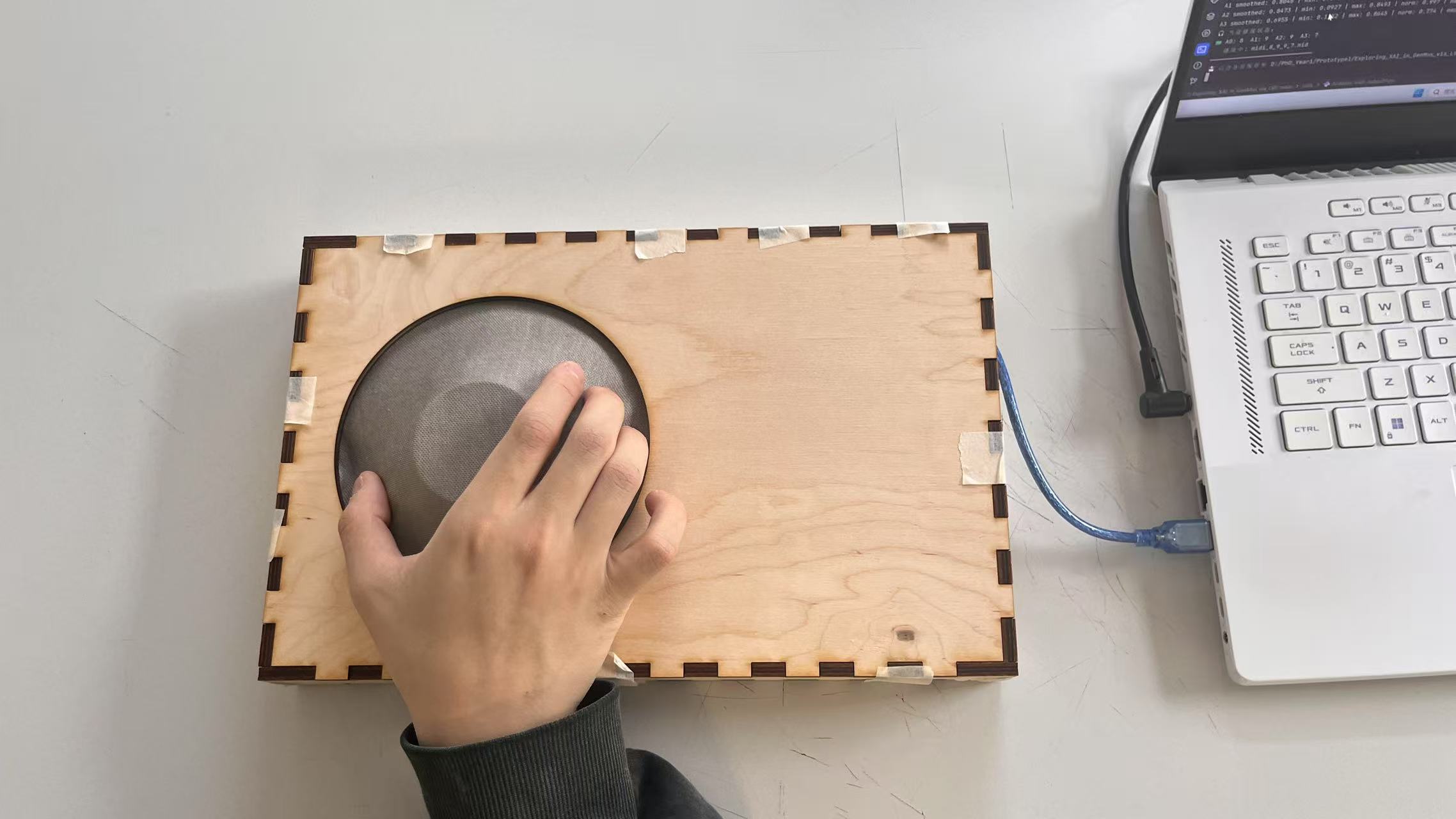}
        \caption{Participants interacted with DeformTune.}\label{fig:fig:deformtune_interaction}
        \Description{Participants interacted with DeformTune.}
    \end{minipage}
\end{figure}

DeformTune combines a deformable tactile interface with a controllable generative model (MeasureVAE \cite{pati_learning_2019}).
The system comprises three components illustrated in Figures \ref{fig:deformtune_structure} and \ref{fig:fig:deformtune_interaction}: i)	A deformable interface made of conductive fabric; ii)	An Arduino-based sensing module with four pressure sensors (using Velostat, copper tape, and sponge). In the Arduino hardware module, the main functions are data processing and transmission. The system applies smoothing to the four sensor readings, using
a continuous linear mapping from pressure data to latent space coordinates, converting fluctuating decimal values(ranging from 0-1) into stable integers(ranging from 1-10). It also continuously updates a dynamic baseline to prevent unintended data transmission when no pressure is applied. 
iii)	A Python backend that processes input from the sensors. In the Python program, the four integer values are received from the Arduino - these correspond to dimensions 0, 1, 2, and 3 of the MeasureVAE latent space, which were assigned to rhythmic complexity, note range, note density, and average interval jump, respectively when using latent space regularisation in training MeasureVAE. These values are then combined in sequence to form the filename of a specific MIDI file to select MIDI 
sequences from pre-generated MIDI files corresponding to a discrete latent vector in the MeasureVAE latent space\cite{bryan-kinns_exploring_2023}, and plays it back in real time. Higher pressure readings result in higher mapped integer values (closer to 10), thus increasing the complexity of the generated MIDI files. 
In this way each sensor maps to one dimenions of MeasureVAE's latent space.

\begin{figure}[htbp]
    \begin{minipage}[t]{0.5\linewidth}
        \centering
        \includegraphics[width=0.5\textwidth]{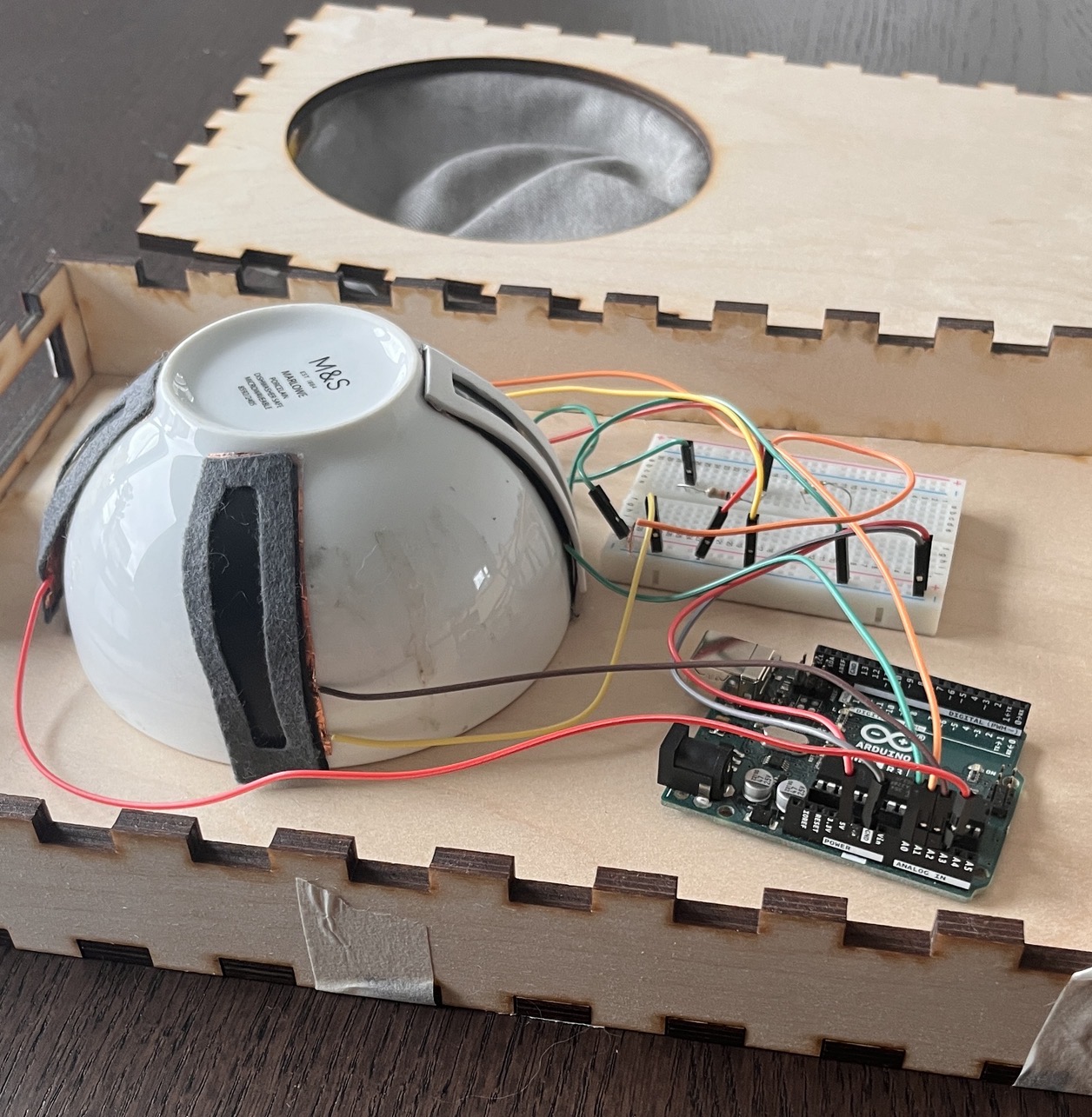}
        \caption{Circuit connections.}\label{fig:circuit connections}
        \Description{To show the structure of DeformTune.}
    \end{minipage}%
    \begin{minipage}[t]{0.5\linewidth}
        \centering
        \includegraphics[width=0.5\textwidth]{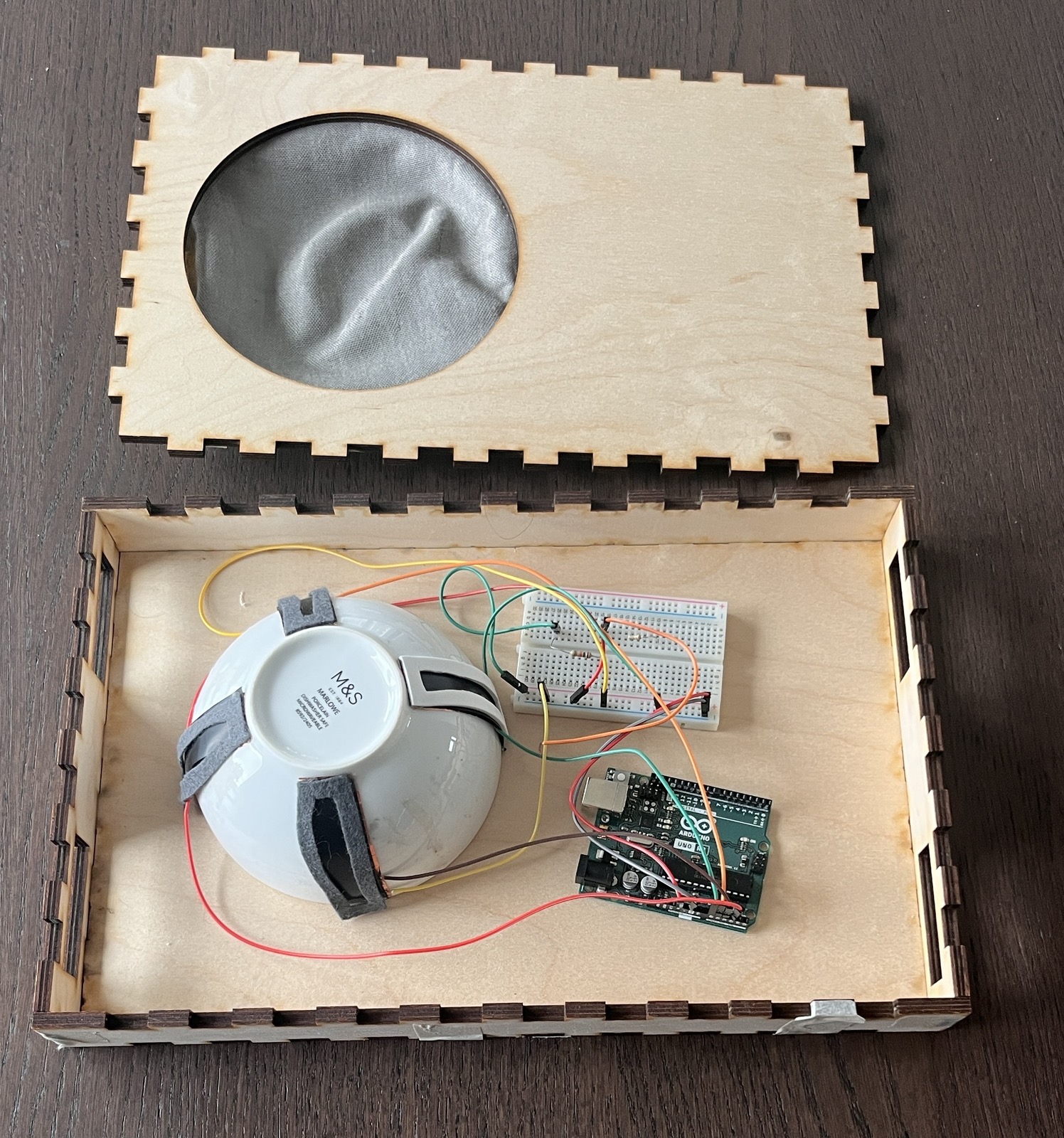}
        \caption{The spatial arrangement of the sensors on the interface.}\label{fig:spactial arrangement}
        \Description{Participants interacted with DeformTune.}
    \end{minipage}
\end{figure}

As shown in Figure \ref{fig:circuit connections}, the sensor structure consists of a ground wire (GND) and a signal transmission line (connected to Arduino analog pins such as A0–A3), both connected to copper tape. The 5V power line is also connected to the conductive fabric which is secured on top of the wood box via copper tape. From bottom to top, the sensor is composed of copper tape, velostat, and foam layers. Figure \ref{fig:spactial arrangement} illustrates the spatial arrangement of the sensors on the interface.

The system supports multi-finger pressing as the primary gesture, which controls the musical parameters of rhythmic complexity, note range, note density, and average interval jump.

The latent space of the MeasureVAE model \cite{bryan-kinns_exploring_2023} in our system was discretised in advance into 10×10×10×10 possible values (each of the four dimensions has 10 pre-generated levels). It precomputed and stored a unique MIDI musical phrases for each of these 10,000 latent combinations. This design sacrifices the ability to generate arbitrary new outputs on the fly (since the system selects from a fixed set of pre-generated samples rather than synthesizing novel ones in real time), but it provides a highly structured and transparent mapping from user input to musical output.

\section{Results}

UEQ \cite{schrepp_design_2017} and RiCE \cite{ford_towards_2023} questionnaire results suggest that DeformTune was enjoyable and expressive, with high Hedonic Quality (M = 1.455) , and strong scores in Creativity Potential (7.42) and Expressivity (7.06). Pragmatic Quality was lower (–0.045), indicating usability challenges, but overall feedback highlighted engaging and creative experiences. For brevity, detailed statistical analysis is omitted in favor of qualitative findings.


Through thematic analysis \cite{braun_using_2006} of interviews with all 11 participants, we identified four key themes related to user interaction experiences and explainability needs when engaging with the DeformTune system.

\textbf{Theme 1: Ambiguity in Action–Sound Mapping.} All 11 participants felt that their physical input had some effect on the music, but 10 of them reported difficulty in understanding exactly what was being controlled. As P2 noted, "I could feel something change, but I couldn’t figure out the pattern". P8 described the system as "confusing to use... I wanted to find a way to control it". Users often described the system as "responsive", but not in a predictable or repeatable way. This uncertainty made it difficult for users to build a reliable mental model of control. To address this issue, participants suggested providing additional feedback during interaction, for instance, LED lights or on-screen icons that activate when a sensor is pressed. 

P5 reported that the conductive fabric covering added a sense of ambiguity. At first, He assumed the interaction method was sliding, when in fact it was pressing. The shape of the sensor led to greater expectations and confusion about how to interact with it. He suggested redesigning the physical shape of sensor to align more closely with everyday gestures such as pressing, sliding, or bending, which could enhance the inherent explainability of the hardware itself.

\textbf{Theme 2: The Tension Between Mystery and Control.} Participants found the interaction experience novel and engaging, partly because of its unpredictability. However, they also expressed frustration when they were unable to form reliable strategies to shape the musical output. P4 said, "I liked a certain sound but couldn’t recreate it". This tension between exploratory play and intentional creation was a recurring theme.
For some users, the lack of a clear interface sparked curiosity and encouraged experimentation. For others, it disrupted their flow and hindered emotional expression. Notably, four participants explicitly stated that they wished for “some rules” to help stabilize their interaction—without fully removing the system’s sense of mystery.
These responses often overlapped, underscoring the need for layered and context-sensitive explainability throughout the user journey.

\textbf{Theme 3: The Need for Multimodal Feedback and Learning Cues.} Nine out of 11 participants reported that haptic input alone was not sufficient to understand how to control the system. They suggested adding visual, auditory, or enhanced tactile feedback—such as pressure indicators, sensor activation cues, or short tutorials—to better support perception and learning.
P8 commented, "Because they’re long, I instinctively wanted to slide, not press". This mismatch created confusion and increased the cognitive load. Several users expressed the need for onboarding support, such as guided demonstrations or explanations of each sensor’s effect, especially for first-time users.
These findings point to the importance of 
clear physical affordances in helping users understand and confidently engage with the system.

\textbf{Theme 4: Explainability as a Means of Creative Empowerment.} A total of five participants expressed interest in integrating their own musical preferences or training data into the system, while two others conceptualized it as a “musical toy” or a “self-reflective instrument”. 
In addition, a total of 9 participants explicitly expressed a desire to control more diverse musical attributes, such as rhythm, melody, harmony, timbre, or pitch. P9 shared, "I hoped to use both hands or control more elements like pitch or rhythm".
Regarding interaction styles, all 11 participants expressed a desire for more diverse interaction modalities, including squeezing, pressing, twisting, sliding, and tapping.

\section{Discussion and Future work}

Our analysis highlights key opportunities to enhance systems like DeformTune by embedding explainability in ways that foster creativity, learning, and emotional connection.

\textbf{Opportunity: Clarifying Action–Sound Mappings through Visual and Tactile Cues.}
One core XAI challenge and opportunity is enabling users to intuitively understand how their input shapes musical output. 
Beyond verbal explanations, interactive overlays (e.g., pressure indicators, mapping diagrams) and haptic feedback can enhance users' understanding of how their physical inputs influence not just sound, but the underlying generative process. For explainable AI systems, it is crucial to clarify how user actions control the behavior of the AI model. To support trust and usability, designs should emphasize the repeatability of AI responses to similar inputs, ensuring that users can reliably learn and predict the effects of their interaction over time.

\textbf{Opportunity: Balancing Mystery and Control through Layered Explanations.}
Fully transparent systems may undermine exploration, while entirely opaque ones can lead to user frustration. Layered XAI design allows for brief, contextual cues during interaction—such as haptic pulses or immediate sonic shifts—to foster smooth, intuitive control, with more detailed post-hoc explanations available after music creation (e.g., “Why did this fragment sound like that?”). This approach helps maintain flow while supporting reflection and mastery.

\textbf{Opportunity: Guided and Gradual Explanation.}
Based on user feedback, one way to lower the learning curve is to include short onboarding animations before first use. Or, a guided exploratory mode embedded within the system that supports step-by-step discovery of action-to-sound mappings. For example, users could focus on one input channel at a time to observe how different gestures affect musical features such as pitch or rhythm.

\textbf{Opportunity: From Technical Transparency to Creative Explainability.}
User feedback revealed that participants primarily sought functional understanding rather than technical transparency. Instead of explaining algorithms or training data, systems can provide explanations grounded in perceptually salient musical features—such as chords, rhythmic complexity, or expressive range. Furthermore, future designs should consider which output attributes are most intuitively perceived by the target user group. For non-experts, parameters such as pitch, loudness, and timbre may be more effective than abstract latent dimensions like note density or average interval jump.

\section{Conclusion}

While the current version of DeformTune remains limited in real-time responsiveness, controllability, and explainability, our user study uncovered critical interaction challenges—such as unclear gesture-to-sound mappings and limited expressive control—as well as concrete user needs for AI explainability. Our thematic analysis reveals promising directions for layered, intuitive, and creatively empowering forms of explainability in AI music systems. These insights can inform the design of future AI music tools that are more transparent, engaging, and usable for novice users.

\bibliographystyle{ACM-Reference-Format}
\bibliography{main}


\end{document}